\documentclass[12pt,reqno]{amsart}
\usepackage[final]{graphicx}
\usepackage{amsfonts}
\usepackage{amsmath}
\usepackage{amssymb}
\usepackage{amsthm}

\newtheorem{theorem}{Theorem}[section]
\newtheorem{lemma}[theorem]{Lemma}
\newtheorem{corollary}[theorem]{Corollary}

\newcommand{\upchi}{\raise1pt\hbox{$\chi$}}
\newcommand{\R}{{\mathord{\mathbb R}}}

\newcommand{\Z}{{\mathord{\mathbb Z}}}

\begin{document}

\title[Minimizing the ground state energy]{Minimizing the ground
state energy of an electron\\ in a randomly deformed lattice}
\author[Baker]{Jeff Baker$^1$}
\author[Loss]{Michael Loss$^2$}
\author[Stolz]{G\"unter Stolz$^1$}

\address{$^1$ University of Alabama at Birmingham, Department of Mathematics, Birmingham AL 35294-1170, sjbaker@uab.edu, stolz@uab.edu}

\address{$^2$ Georgia Institute of Technology, School of Mathematics,
Atlanta GA 30332-0160,\\ loss@math.gatech.edu}

\date{\today}
\maketitle

\vspace{.3truein}
\centerline{\bf Abstract}

\medskip
{\sl We provide a characterization of the spectral minimum for a
random Schr$\ddot{\mbox{o}}$dinger operator of the form $H=-\Delta
+ \sum_{i \in \Z^d}q(x-i-\omega_i)$ in $L^2(\R^d)$, where the
single site potential $q$ is reflection symmetric, compactly
supported in the unit cube centered at $0$, and the displacement
parameters $\omega_i$ are restricted so that adjacent single site
potentials do not overlap. In particular, we show that a
minimizing configuration of the displacements is given by a
periodic pattern of densest possible $2^d$-clusters of single site
potentials.

The main tool to prove this is a quite general phenomenon in the
spectral theory of Neumann problems, which we dub ``bubbles tend
to the boundary.'' How should a given compactly supported
potential be placed into a bounded domain so as to minimize or
maximize the first Neumann eigenvalue of the Schr\"odinger
operator on this domain? For square or rectangular domains and
reflection symmetric potentials, we show that the first Neumann
eigenvalue is minimized when the potential sits in one of the
corners of the domain and is maximized when it sits in the center
of the domain. With different methods we also show a corresponding
result for smooth strictly convex domains. }

\section{Introduction and Main Results}

\subsection{The Displacement Model} \label{subsec:displace}

The one electron model of solid state physics describes the
behavior of a single electron moving under the presence of an
exterior force generated by the effective potentials of a fixed
configuration of nuclei in a solid. Also disregarding
electron-electron interactions, this results in the one-electron
Schr\"odinger operator
\[
H = -\Delta + V
\]
in $L^2(\R^d)$, where
$-\Delta$ and $V$ are the kinetic and potential energy of the
particle, respectively.

Typically one chooses a potential $V$ that effectively models the
characteristics of a particular solid. For example, one might use a
periodic potential to model a crystal or other well ordered
substance. While for a material containing a sufficient number of
impurities, or disorder, one might use a random potential. In this
paper we consider a potential generated by identical atoms or ions
located at the points $i+\omega_i$, $i\in \Z^d$, i.e.\
\[
V_{\omega}(x) = \sum_{i \in \Z^d} q(x-i-\omega_i).
\]
We refer to $q$ as the \emph{single site potential} and consider
real valued single site potentials $q \in L^{\infty}$ which are
reflection symmetric, i.e.\ symmetric in each variable with the
remaining variables fixed, and compactly supported in the unit
cube $\Lambda_0 := (-\frac{1}{2},\frac{1}{2})^d$ of $\R^d$, i.e.
supp $q \subset [-r,r]^d \subset \Lambda_0$, $r<1/2$. We denote
the collection of displacements by $\omega = \{\omega_i\}_{i \in
\Z^d}$, where each $\omega_i \in [-d_{max},d_{max}]^d$. Finally we
choose $r+d_{max} = \frac{1}{2}$, which insures that adjacent
single site potentials in the sum above do not overlap.

For any possible collection of displacements $\omega$,
\begin{equation} \label{lbl:displacementmodel}
H(\omega) := -\Delta + V_{\omega}
\end{equation}
with domain $H^2(\R^d)$, the second order Sobolev space, defines a
self adjoint operator in $L^2(\R^d)$. We will refer to the family
$H(\omega)$ as the \emph{displacement model}. As $V$ is uniformly
bounded with respect to $\omega$, the spectrum of $H(\omega)$,
$\sigma(H(\omega))$, is uniformly bounded from below.

The question we will address is the following: How can one
characterize
\begin{equation} \label{eqn:groundstate}
E_0 := \inf_{\omega} \inf \sigma(H(\omega)),
\end{equation}
i.e.\ the infimum of the ground state energy of $H(\omega)$ for all
possible nuclear configurations $\omega$? In particular, is there a
minimizing configuration $\omega^{min}$ such that
\begin{equation} \label{eqn:minimizer}
\inf \sigma(H(\omega^{min})) = E_0,
\end{equation}
and how does it look like?

Our main result is the answer to this question:

\begin{theorem} \label{thm:main}
A minimizing configuration $\omega^{min}$ for the ground state
energy of $H(\omega)$ in the sense of \eqref{eqn:minimizer} is
given by
\begin{equation} \label{eqn:clusters}
\omega_i^{\min} = ((-1)^{i_1} d_{max}, \ldots, (-1)^{i_d} d_{max})
\end{equation}
for all $i = (i_1,\ldots,i_d) \in \Z^d$.
\end{theorem}

\begin{figure}[h]
  \centering
  \includegraphics[width=0.35\textwidth]{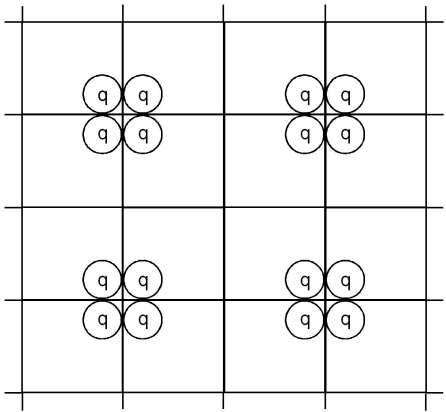}
  \caption{The support of $V_{\omega^{min}}$ for $d=2$ and radially
symmetric $f$.} \label{fig:minimizer}
\end{figure}

The energy minimizing potential $V_{\omega^{min}}$ is $2$-periodic
in each coordinate and given by the densest possible cluster of the
nuclei in the period cell $(-\frac{1}{2}, \frac{3}{2})^d$, namely
all single site potentials within the cluster move as close to the
center $(\frac{1}{2},\ldots, \frac{1}{2})$ of the period cell as
possible, see Figure~\ref{fig:minimizer}.
Thus, by Floquet-Bloch theory \cite{Barry4}, $E_0$ is the lowest
eigenvalue of $-\Delta + V_{\omega^{min}}$ restricted to
$L^2((-\frac{1}{2}, \frac{3}{2})^d)$ with periodic boundary
conditions. Due to the symmetry of the $2^d$-cluster this is the
same as the ground state energy of $-\Delta +q(x-(d_{max},\ldots,
d_{max}))$ on $L^2(\Lambda_0)$ with Neumann boundary conditions.
For $d=1$ Theorem~\ref{thm:main} has been conjectured and
partially proven in \cite{lott}.

Our interest in this result is mostly motivated by the case where
the displacements $\omega_i$, $i\in \Z^d$, are independent,
identically distributed $\R^d$-valued random variables with
$[-d_{max},d_{max}]^d$ as the support of their common distribution $\mu$.
In this case we refer to $H(\omega)$ as the \emph{random
displacement model}, which is ergodic with respect to shifts in
$\Z^d$. Thus its spectrum is almost surely deterministic, i.e.\
there exists $\Sigma \subset \R$ such that
\[
\sigma(H(\omega)) = \Sigma \quad \mbox{for a.e.\ $\omega$},
\]
e.g.\ \cite{Carmona/Lacroix}.

Compared to other prominent models of random Schr\"odinger
operators, e.g.\ the Anderson model or Poisson model, few rigorous
results are known for the random displacement model. This is
mostly due to the fact that $H(\omega)$ does not depend
monotonically (in form sense) on the random parameters $\omega_i$.
Even the structure of the almost sure spectrum $\Sigma$ is
unclear. It can be said that
\begin{equation} \label{eq:periodicsupport}
\Sigma = \overline{ \bigcup_{\omega \in C_{per}} \sigma(H(\omega))},
\end{equation}
where $C_{per}$ is the set of all configurations $\omega: \Z^d \to
\mbox{supp}\,\mu$ which are periodic with respect to some sublattice
of $\Z^d$. This follows by adapting the proof of the corresponding
result for Anderson models, given e.g.\ in \cite{kirsch}.

A consequence of Theorem~\ref{thm:main} is
\begin{corollary} \label{cor:infsigma}
The infimum of the almost sure spectrum of the random displacement
model is given by
\[
\inf \Sigma = E_0 = \inf \sigma(H(\omega^{min})).
\]
\end{corollary}

\begin{proof} This follows from \eqref{eq:periodicsupport} since
$\omega^{min} \in C_{per}$ and, by Theorem~\ref{thm:main}, $\inf
\sigma(H(\omega^{min})) \le \inf \sigma(H(\omega))$ for all other
$\omega\in C_{per}.$
\end{proof}

Note that, at least for sign-definite $q$, the answer to the same
question for the Anderson or Poisson model is quite
straightforward and found by considerations involving not much
more than minimizing the potential energy: For the Anderson model,
$\inf \Sigma$ is found by choosing all random couplings minimal.
For the Poisson model one has $\inf \Sigma=0$ if $q\ge 0$ and
$\inf \Sigma=-\infty$ if $q\le 0$. In fact, the latter, with few
exceptions, only requires that the negative part of $q$ doesn't
vanish \cite{Andoetal}. For the Anderson model with
sign-indefinite $q$ the description of $\Sigma$ or just $\inf
\Sigma$ causes difficulties similar to those for the random
displacement model. Najar \cite{najar} has a result for this case
in the small coupling regime, proven by perturbative arguments.

What makes our result, as well as the techniques in its proof,
rather interesting is that minimizing the spectrum in the
displacement model requires an understanding of the interaction
between kinetic and potential energy. Physically, one can understand
our result best for the case of negative potential wells $q$. In
this case the formation of clusters of $2^d$ sites allows for states
with low potential energy without sacrificing much kinetic energy.
But we stress that Theorem~\ref{thm:main} and
Corollary~\ref{cor:infsigma} hold without any sign-restriction on
$q$.

For the multi-dimensional random displacement model it is not yet
known if the spectrum is localized, in the sense of being pure
point, near the bottom of the spectrum. This is in contrast to the
situation for Anderson and Poisson models. For the Anderson model
this is a long standing result, with the hardest case of Bernoulli
distributed random couplings recently settled in
\cite{bourgain/kenig}. The new type of multi scale analysis
introduced in \cite{bourgain/kenig} was now also used to prove the
corresponding fact for the Poisson model in arbitrary dimension
\cite{GHK1,GHK2}. For the one-dimensional displacement model,
localization at all energies was proven in \cite{buschmann/stolz}
and, with different methods and under more general assumptions, in
\cite{damanik/sims/stolz}. The only available result on
localization for the multi-dimensional displacement model is
Klopp's work \cite{klopp}, establishing the existence of a
localized region for the semiclassical version $-h^2\Delta
+V_{\omega}$ of \eqref{lbl:displacementmodel} if $h$ is
sufficiently small.

Theorem~\ref{thm:main} should serve as a first step towards
understanding the spectral type of $-\Delta+V_{\omega}$ near $\inf
\Sigma$ by identifying the periodic configuration in which $\inf
\Sigma$ is attained. An important further step towards
localization would be to quantify probabilistically how many other
configurations have ground states close to $\inf\Sigma$, that is,
to prove smallness of the integrated density of states (IDS) near
$\inf\Sigma$ (or a related finite volume property). To this end it
is interesting to note that for $d=1$ the configuration given by
\eqref{eqn:clusters} (in this case ``dimerization'') is only one
of many minimizing periodic configurations. This will have
interesting consequences for the IDS. In particular, one may not
find the Lifshitz tail behavior familiar from Anderson and Poisson
models and the exact asymptotics may depend strongly on the
displacement distribution $\mu$. However, we believe that, under
suitable assumptions on $q$ suggested by our proof of
Theorem~\ref{thm:main}, in $d\ge 2$ the configuration
\eqref{eqn:clusters} is the unique periodic minimizer. We plan to
investigate this further in a separate work.

\subsection{Bubbles tend to the boundary} \label{subsec:bubbles}

Theorem~\ref{thm:main} amounts to optimizing the infinitely many
parameters $\omega_i$, $i\in\Z^d$, with respect to minimizing the
spectrum. Surprisingly, as will be shown in
Section~\ref{sec:clustering}, its proof can be reduced to the
following spectral optimization result in just one parameter.

\begin{theorem} \label{thm:neumannsquare}
Let $q$ be as above, i.e.\ bounded, reflection symmetric, and
supported in $[-r,r]^d$ for some $r<\frac{1}{2}$. Let
$d_{max}=\frac{1}{2}-r$ and for $a\in [-d_{max},d_{max}]^d$ let
$H_{\Lambda_0}^N(a) := -\Delta+q(x-a)$ in $L^2(\Lambda_0)$ with
Neumann boundary conditions on $\partial \Lambda_0$ and denote the
ground state energy of $H_{\Lambda_0}^N(a)$ by $E_0(a)$. Then we
have the following alternative: Either

\begin{itemize}
\item[(i)] $E_0(a)$ is strictly maximized at $a=0$ and strictly
minimized in the $2^d$ corners\\ $(\pm d_{max}, \ldots, \pm
d_{max})$ of $[-d_{max},d_{max}]^d$
\end{itemize}
or
\begin{itemize}
\item[(ii)] $E_0(a)$ is identically zero. In this case the
corresponding eigenfunction is constant outside of the support of
$q$.
\end{itemize}
\end{theorem}

In fact, in will show that in case (i) the function $E_0(a)$ is
partially strictly decreasing away from the origin, i.e.\ that
whenever all but one of the variables $(a_1,\ldots,a_d)$ are
fixed, then $E(a_1,\ldots,a_d)$ is strictly decreasing for the
remaining variable in $[0,1/2-r]$ and, by symmetry, strictly
increasing in $[-1/2+r,0]$.

A sufficient, but far from necessary condition for case (i) to
hold is that $q$ has fixed sign and does not vanish identically,
as in this case $E_0(a)$ never vanishes. Case (ii) happens if the
Neumann problem for $-\Delta +q$ on the
support of $q$ has lowest eigenvalue $0$. Non-vanishing $q$ with
this property are easily constructed.

We find Theorem~\ref{thm:neumannsquare} quite interesting for its
own sake, independent of its application to prove
Theorem~\ref{thm:main}. It is a prototype of what seems to be a
very general phenomenon appearing for Neumann problems on bounded
domains, namely that ``bubbles tend to the boundary''. To this
end, we have the following result for general strictly convex
smooth domains and smooth potentials, proven in
Section~\ref{sec:neumannconvex} with a method very different from
the one we use in Section~\ref{sec:neumannsquare} to prove
Theorem~\ref{thm:neumannsquare}.

Consider an open, bounded domain $D \subset \R^d$ with smooth
boundary. We shall assume that $D$ is \emph{strictly convex}. Let
$q(x)$ be any bounded smooth potential whose support is a subset
of $D$. For $a\in \R^d$ let $q_a(x):= q(x-a)$. In $D$ consider the
Schr\"odinger operator
\[
H_D^N(a) = -\Delta + q_a
\]
with Neumann boundary conditions on $\partial D$ (where
restriction of $q_a$ to $D$ is implied). We denote its ground
state energy by $E_0(a)$. As shown in Lemma~\ref{lem:conteigval}
of Section~\ref{sec:prelim}, $E_0(a)$ is continuous in $a$.

Denote by $G \subset \R^d$ the collection of vectors $a$ such that
$q_a$ has its support also in $D$. Note that $G$ is an open set.

\begin{theorem}[\bf{Strong minimum principle for $E_0$}]
\label{thm:neumannconvex} If $E_0(a_0) = \inf_{a \in G} E_0(a)$
for some $a_0 \in G$, then $E_0(a)$ is identically zero. In this
case the wave function is constant outside the support of the
potential. In other words if $E_0(a)$ does not vanish identically
in $G$, then $E_0(a_0) > \inf_{a \in G} E_0(a)$ for all $a_0 \in
G$.
\end{theorem}

The continuous function $E_0:\overline{G}\to \R$ must assume its
minimum. By Theorem~\ref{thm:neumannconvex}, if $E_0(a)$ does not
vanish identically, the minimum must be assumed on $\partial G$.
In the same situation Theorem~\ref{thm:neumannsquare} gives the
more precise result that the minimum is assumed in the corners of
$\overline{G} = [-d_{max},d_{max}]^d$.

For radially symmetric $q$ and various types of domains $D$, the
question of minimizing the first \emph{Dirichlet}-eigenvalue of
$-\Delta+q_a$ on $D$ is well studied, see \cite{hersch} for disks
and regular polygons, or \cite{harrell} for a more general class
of domains which have a certain reflection property with respect
to the symmetry axes of the potential.

Common to all results for Dirichlet problems is that the
maximizing and minimizing positions depend on the sign of the
potential. For an \emph{obstacle} $q\ge 0$ (or ``$q=\infty$'',
meaning a hole in the domain marked by an additional Dirichlet
boundary condition) the maximizing position is in the ``center''
of the domain, while the first eigenvalue is minimized when the
obstacle is in contact with the boundary. The reverse is true for
the case of a well $q\le 0$. As pointed out in \cite{harrell},
this is most easily understood, if not proven, by a perturbative
argument: Consider $-\Delta+\lambda q$ on $D$ with Dirchlet
boundary condition. Its lowest eigenvalue $E_0(\lambda)$ satisfies
\begin{equation} \label{eq:feynmanhellmann}
E_0'(0)= \int_D q|u_0|^2\,dx,
\end{equation}
where $u_0$ is the ground state eigenfunction of the Dirichlet
Laplacian. Thus $E_0(\lambda)$ changes the most (least), if $q$ is
placed where $u_0$ is largest (smallest), which is near the center
(boundary) of $D$. The sign of $q$ determines the sign of $E_0'$
and thus reverses the role of maximizer and minimizer.

This motivation through first order perturbation theory fails for
the Neumann problem. In this case \eqref{eq:feynmanhellmann} still
applies, but the ground state of the Neumann Laplacian is constant
and thus $E_0'(0)$ is independent of the placement of $q$. This
explains why the Neumann version of the problem is more subtle
than the Dirichlet problem (roughly by one order of perturbation
theory). Consequently, the methods used in \cite{hersch} and
\cite{harrell} do not extend to give similar results for the
Neumann case. An exception is a remark in \cite{hersch} concerning
infinite spherical obstacles in spherical domains. The only other
work on the Neumann case, which we found in the literature, is
\cite{koloko}, which gives perturbative and numerical results
concerning the optimal configurations of small Dirichlet holes in
planar domains for maximizing the first Neumann eigenvalue.

We indeed use a second order perturbation theory formula as the
starting point of the proof of Theorem~\ref{thm:neumannconvex},
see \eqref{eq:2ndorderpert} below. Our proof of
Theorem~\ref{thm:neumannsquare} doesn't use perturbation theory
(but Floquet-Bloch theory, and the variational characterization of
ground states, convexity of the kinetic energy, and unique
continuation of harmonic functions). Still, the result may be
motivated by second order perturbation theory:

In $d=2$ (for simplicity) consider the Neumann problem
$-\Delta+\lambda q$ on $L^2((-\frac{1}{2},\frac{1}{2})^2,dxdy)$.
The lowest eigenvalue $E_0(\lambda)$ satisfies the second order
perturbation formula
\begin{equation} \label{eq:secondorder}
E_0''(0) = -2\sum_{k>0} \frac{(u_0,qu_k)^2}{E_k-E_0},
\end{equation}
where $E_k$ and $u_k$ are the higher eigenvalues and
eigenfunctions of the Neumann-Laplacian, see
Section~\ref{subsec:pert}. Considering only the leading term of
\eqref{eq:secondorder}, corresponding to $E_1=E_2=\pi^2$, we get
\[ -\frac{4}{\pi^2} \left[ \left(\int q(x,y)\sin(\pi x)\,dx\,dy\right)^2 +
\left(\int q(x,y)\sin(\pi y)\,dx\,dy\right)^2 \right], \] which is
negative, independent of the sign of $q$. If $q=q_a$, with $q_0$
reflection symmetric and of fixed sign, then both integrals are
zero for $a=0$, and both integrals become maximal (in absolute
value) if $a$ is located near one of the four corners
$(\pm\frac{1}{2},\pm\frac{1}{2})$. Again, this is independent of
the sign of $q$.

\section{Preliminaries} \label{sec:prelim}

Throughout this section $\Omega \subset \R^d$ will be open and
bounded. The Neumann Laplacian $-\Delta_{\Omega}^N$ on $\Omega$ is
the unique selfadjoint operator whose quadratic form is
\[ \int_{\Omega} |\nabla f(x)|^2\,dx \]
for $f$ in the domain $H^1(\Omega)$, the first order Sobolev
space.

\subsection{Continuity of Eigenvalues} \label{subsec:cont}

Assume that $\Omega$ satisfies the $H^1$-extension property, i.e.\
there exists a bounded operator $E:H^1(\Omega)\to H^1(\R^d)$ such
that $(Ef)(x) = f(x)$ for all $f\in H^1(\Omega)$ and almost every
$x\in \Omega$. Note that a sufficient condition for this is that
$\Omega$ has Lipschitz boundary, e.g.\ Theorem~V.4.12 of
\cite{Edmunds/Evans}.

Let $q\in L^{\infty}(\R^d)$ be real-valued and define $q_a(x) =
q(x-a)$ for all $a\in \R^d$. Let $H_{\Omega}^N(a) =
-\Delta_{\Omega}^N+q_a$.

\begin{lemma} \label{lem:conteigval}
$H_{\Omega}^N(a)$ has purely discrete spectrum consisting of
eigenvalues $E_0(a) \le E_1(a) \le \ldots \le E_n(a)$, counted
with multiplicity, where all functions $E_n$ are continuous in
$a$.
\end{lemma}

\begin{proof}
Fix $C> \|q\|_{\infty}$. From the extension property of $\Omega$
and boundedness of $q$ it follows that
$(-\Delta_{\Omega}^N+C)^{-1}$ and $(H_{\Omega}^N(a)+C)^{-1}$ are
compact, e.g.\ Theorem~V.4.13 of \cite{Edmunds/Evans}. Thus
$H_{\Omega}^N(a)$ has purely discrete spectrum $E_0(a) \le E_1(a)
\le \ldots$. It remains to show that $(H_{\Omega}^N(a)+C)^{-1}$ is
norm-continuous in $a$. Continuity of the eigenvalues of
$(H_{\Omega}^N(a)+C)^{-1}$, and thus the eigenvalues of
$H_{\Omega}^N(a)$, then follows from the min-max-characterization
of eigenvalues.

Without restriction, consider continuity at $a=0$. For any $p\in
(d,\infty)$ with $p\ge 2$ one has, e.g.\ Theorem~4.1 of
\cite{SimonTrace},
\begin{eqnarray} \label{eq:tracebound}
\|\chi_{\Omega}(q_a-q) (-\Delta+C)^{-1/2} \| & \le & (2\pi)^{-d/p}
\|\chi_{\Omega}(q_a-q)\|_p \|(|\cdot|^2+C)^{-1/2}\|_p \nonumber \\
& \rightarrow & 0 \quad \mbox{as $a\to 0$}.
\end{eqnarray}

As $E:H^1(\Omega)\to H^1(\R^d)$ is bounded, it follows that
$(-\Delta+C)^{1/2} E (-\Delta_{\Omega}^N+C)^{-1/2}$ is bounded
from $L^2(\Omega)$ to $L^2(\R^d)$. Combined with
\eqref{eq:tracebound} this yields
\begin{equation} \label{eq:relbound}
\|\chi_{\Omega} (q_a-q) (-\Delta_{\Omega}^N+C)^{-1/2}\| \to 0
\quad \mbox{as $a\to 0$}.
\end{equation}

Norm-continuity of $(H_{\Omega}^N(a)+C)^{-1}$ at $a=0$ now follows
from \eqref{eq:relbound}, boundedness of
$(-\Delta_{\Omega}^N+C)^{1/2}(H_{\Omega}^N(0)+C)^{-1/2}$ and the
resolvent identity
\[ (H_{\Omega}^N(a)+C)^{-1} -(H_{\Omega}^N(0)+C)^{-1} = (H_{\Omega}^N(a)+C)^{-1} \chi_{\Omega} (q_a-q)
(H_{\Omega}^N(0)+C)^{-1}. \]
\end{proof}

\subsection{Positivity and non-degeneracy of the ground state}
\label{subsec:pos}

We will frequently use that for the domains considered by us and
bounded potentials $q$ the ground state energy $E_0$ of
$H=-\Delta_{\Omega}^N+q$ is non-degenerate and that the
corresponding eigenfunction can be chosen strictly positive. This
generally holds if, in addition to the assumptions from
Section~\ref{subsec:cont}, $\Omega$ is connected. The latter
guarantees that the ground state energy $0$ of
$-\Delta_{\Omega}^N$ is non-degenerate ($-\Delta_{\Omega}^N
\varphi=0$ implies that $\int_{\Omega} |\nabla \varphi|^2\,dx=0$,
i.e.\ $\nabla \varphi\equiv 0$ and thus $\varphi$ constant by
connectedness). Non-degeneracy and positivity of the ground state
of $H$ follows from the general theory of positivity preserving
operators provided in Section~XIII.12 and the following Appendix 1
of \cite{Barry4}.

\subsection{Perturbation formulas} \label{subsec:pert}

For completeness, let us briefly recall the derivation of the
eigenvalue perturbation formulas which we use in our arguments.
Most significantly, this will be the first and second order
perturbation formulas \eqref{eq:firstorderpert} and
\eqref{eq:2ndorderpert} with respect to displacements of the
potential in Section~\ref{sec:neumannconvex}. However, they follow
in the same way as the corresponding formulas for coupling
constant dependence, e.g.\ \eqref{eq:secondorder}, so we will
focus on the latter.

Let $\Omega$ and $q$ satisfy the assumptions of the previous two
subsections, $H(\lambda) = -\Delta_{\Omega}^N+\lambda q$, $E_k =
E_k(\lambda)$ its eigenvalues ordered by $E_0 < E_1\le E_2
\le\ldots$ and $u_k = u_k(\cdot, \lambda)$ corresponding real
normalized eigenfunctions. Then
\begin{equation} \label{eq:lambda1storder}
E_0'(\lambda) = (u_0,qu_0)
\end{equation}
and
\begin{equation} \label{eq:lambda2ndorder}
E_0''(\lambda) = -2 \sum_{k>0} \frac{(u_0,qu_k)^2}{E_k-E_0}\,.
\end{equation}
The formulas \eqref{eq:firstorderpert} and \eqref{eq:2ndorderpert}
below follow with the same argument, using smoothness of $q$ and
differentiating separately with respect to each component of $a$
(the extra term $(u_0,(\Delta q_0)u_0)$ in \eqref{eq:2ndorderpert}
does not appear in \eqref{eq:lambda2ndorder} as
$\partial_{\lambda}^2(\lambda q)=0$).

The equation \eqref{eq:lambda1storder} is the classical
Feynman-Hellmann formula, derived by using non-de\-ge\-ne\-racy of
$E_0$ (and thus analyticity of $E_0$ and $u_0$ in $\lambda$) and
the fact that $(u_0, \partial_{\lambda} u_0)=0$. Differentiating
\eqref{eq:lambda1storder} and using completeness of the $u_k$ we
get
\begin{eqnarray} \label{eq:eigfunctexp}
E_0''(\lambda) & = & 2(\partial_{\lambda} u_0, qu_0) \nonumber \\
& = & 2 \sum_{k>0} (qu_0,u_k)(\partial_{\lambda} u_0,u_k),
\end{eqnarray}
noticing that the $k=0$ term vanishes. Differentiating the
eigenvalue equation $-\Delta u_0 +\lambda q u_0= E_0 u_0$ yields,
for every $k>0$,
\[ qu_0 = (E_0-E_k)\partial_{\lambda} u_0 +(u_0,qu_0)u_0- (-\Delta+\lambda
q-E_k) \partial_{\lambda} u_0, \] and thus
\begin{equation} \label{eq:lambdaderiv}
\partial_{\lambda} u_0 = \frac{qu_0 -(u_0,qu_0)u_0 + (-\Delta +\lambda
q-E_k)\partial_{\lambda} u_0}{E_0-E_k} \, .
\end{equation}
After noting that $((-\Delta+ \lambda q-E_k) \partial_{\lambda}
u_0, u_k)=0$, \eqref{eq:lambda2ndorder} follows from inserting
\eqref{eq:lambdaderiv} into \eqref{eq:eigfunctexp}.

\section{Theorem~\ref{thm:neumannsquare} implies Theorem~\ref{thm:main}}
\label{sec:clustering}

Theorem~\ref{thm:neumannsquare} says that $\inf_a E_0(a) =
E_0(a^{min})$, where $a^{min}$ corresponds to one of the $2^d$
corners of the the cube $[-d_{max},d_{max}]^d$, say $a^{min} :=
(d_{max}, \ldots, d_{max})$. Once we know this, then the central
ideas of the proof of Theorem~\ref{thm:main} are (i) Neumann
bracketing to go from $H(\omega)$ to operators of the type
$H_{\Lambda_0}^N(a)$ and (ii) extending the ground state of the
minimizer $H_{\Lambda_0}^N(a^{min})$ to $\R^d$ by repeated
reflection.

\begin{proof} (of Theorem \ref{thm:main})
For any given configuration $\omega$, the restriction of
$H(\omega)$ to the unit cube centered at $i\in \Z^d$ with Neumann
boundary conditions is unitarily equivalent (via translation by
$i$) to $H_{\Lambda_0}^N(\omega_i)$, defined as in
Theorem~\ref{thm:neumannsquare}. Thus, by Neumann bracketing and
Theorem~\ref{thm:neumannsquare},
\begin{eqnarray*}
\inf \sigma( H(\omega) ) &\geq& \inf \sigma \left(
\bigoplus_{i \in \Z^D} H^N_{\Lambda_0}(\omega_i) \right) \\
&\geq& \inf \left\{E_0(a): \,a\in [-d_{max},d_{max}]^d\right\} \\
&=& E_0(a^{min}).
\end{eqnarray*}
This holds for arbitrary configurations $\omega$ and thus, by
\eqref{eqn:groundstate}, $E_0 \ge E_0(a^{min})$.

Now consider $\omega^{min} = (\omega_i^{min})_{i\in \Z^d}$ as
given by \eqref{eqn:clusters}. The corresponding potential
\[ V_{\omega^{min}}(x) = \sum_{i\in \Z^d} q(x-i-\omega_i^{min}) \]
is $2$-periodic in $x_i$ for each $i$. By Floquet-Bloch theory
\cite{Barry4} the bottom of the spectrum of $H(\omega^{min}) =
-\Delta + V_{\omega^{min}}$ is given by the smallest eigenvalue
$E_0^{per}$ of its restriction to $\Lambda_0^2 := (-\frac{1}{2},
\frac{3}{2})^d$ with periodic boundary conditions, see
Figure~\ref{fig:periodtwomin}.

\begin{figure}[h]
  \centering
  \includegraphics[width=0.25\textwidth]{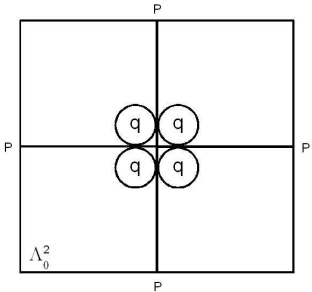}
  \caption{The period cell of $V_{\omega^{min}}$ in $d=2$.}
  \label{fig:periodtwomin}
\end{figure}

On $\Lambda_0^2$ the potential $V^{\omega^{min}}$ is symmetric
with respect to all hyperplanes $x_i=1/2$, $i=1,\ldots,d$. Thus
$E_0^{per}$ coincides with the smallest eigenvalue of the Neumann
problem on $\Lambda_0^2$. Again by symmetry of the potential, the
latter coincides with the smallest eigenvalue of the Neumann
problem on $\Lambda_0$. As $\omega_0^{min} = (d_{max},\ldots,
d_{max}) = a^{min}$, this eigenvalue is $E_0(a^{min})$. In summary
we have shown that
\[ E_0 \le \inf \sigma(H(\omega^{min})) = E_0(a^{min}) \le E_0. \]
Thus $E_0 = \inf \sigma (H(\omega^{min}))$, which proves
Theorem~\ref{thm:main}.
\end{proof}

\section{Proof of Theorem~\ref{thm:neumannsquare}} \label{sec:neumannsquare}

This entire section is devoted to prove
Theorem~\ref{thm:neumannsquare}. Thus we work under the assumptions
that $q\in L^{\infty}$ is real-valued, non-vanishing, reflection
symmetric and supported in $[-r,r]^d$, $r<1/2$.

Suppose that alternative (ii) of Theorem~\ref{thm:neumannsquare} is
false. We will show that this implies that alternative (i) must
hold.

We begin by fixing all of the components of the displacement
parameter except for one, which may be chosen to be the first, and
consider the lowest Neumann eigenvalue as a function only of the
first coordinate, i.e. $E_0(a) := E_0(a,a_2,\ldots,a_d)$. We note
that $E_0(a)$ depends continuously on $a$ (see Lemma
\ref{lem:conteigval}) and that by symmetry we have
\[ E_0(-a) = E_0(a). \]
For this reason we will restrict ourselves to the case $a \in
[0,\frac{1}{2}-r]$ and show that
\begin{equation} \label{eq:Emon}
E_0(a_1) > E_0(a_2) \quad \mbox{for $0 \le a_1 < a_2 \le 1-r$}.
\end{equation}

As the same holds for $E_0$ as a function of each other coordinate,
we conclude from this that $E_0$ has a strict maximum at the origin
and strict minima at the corners $(\pm d_{max},\ldots, \pm
d_{max})$, i.e.\ we are in the situation of alternative (i).

As $(a_2,\ldots,a_d)$ will be kept fixed, we will use the
(slightly sloppy) abbreviation $q(x) := q(x_1,x_2-a_2,\ldots,
x_d-a_d)$ for the rest of the section. For a scalar $a$ and
$e_1=(1,0,\ldots,0)$ we then write $q_a(x) = q(x-ae_1)$, a
notation to be used also for functions other than $q$. By
$\partial_n$ we will denote the exterior normal derivative on the
boundary of a given domain.

\begin{lemma}
$E_0(0) > E_0(a)$ for every $a \in [0, \frac{1}{2}-r]$.
\label{lem:lem1}
\end{lemma}

\begin{proof}
Define the tube $L := \{x : |x_i| < \frac{1}{2}, i=2,\ldots,d \}$
and construct a periodic extension, $q_{per}(x)$, of the potential
$q(x)$ on $L$ by
\[
q_{per}(x) = \sum_{i \in \Z} q_i(x).
\]
\begin{figure}[h]
  \centering
  \includegraphics[width=0.40\textwidth]{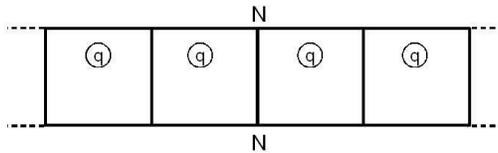}
  \caption{Periodic extension, $q_{per}(x)$, of
  $q(x)$} \label{figperiodone}
\end{figure}
We consider the Neumann problem on $L$ for $-\Delta +
q_{per}(x)$, see Figure~\ref{figperiodone}. Let $\psi$
denote the normalized Neumann ground state of $-\Delta + q(x)$ on
the unit cube $\Lambda_0$. Construct a new function $\Psi$ on the
tube $L$ by periodically extending $\psi$ on all of $L$. Symmetry of
the potential implies that $\Psi$ is a smooth solution of
\begin{eqnarray}
-\Delta \Psi(x) + q_{per}(x)\psi(x) = E_0(0) \Psi(x) \label{eqn:layerproperty}
\end{eqnarray} on all of $L$. Now multiply (\ref{eqn:layerproperty}) by
$\Psi(x)$. Then over any unit cell $C$ in $L$, periodicity
of $\Psi$ implies we may integrate by parts without creating
boundary terms. In particular, this holds for the unit cell
$\Lambda_a := \Lambda_0-a e_1$, yielding
\begin{eqnarray}
E_0(0) &=& \int_{\Lambda_a} \left(|\nabla \Psi(x)|^2 + q_{per}(x) \Psi^2(x)\right) dx
\label{eqn:layerproperty2}.
\end{eqnarray}
Shifting to the right by $a$ does not affect the result of
equation (\ref{eqn:layerproperty2}), i.e.
\begin{eqnarray}
E_0(0) &=& \int_{\Lambda_0} \left(|\nabla \Psi_{a}(x)|^2 + q_{per}(x-a e_1) \Psi_{a}^2(x)\right) dx \nonumber \\
&=& \int_{\Lambda_0} \left(|\nabla \Psi_{a}(x)|^2 + q_{a}(x) \Psi_{a}^2(x)\right) dx.
\label{eqn:layerproperty3}
\end{eqnarray}
While $\Psi_{a}(x)$ does not satisfy Neumann boundary conditions
on $\Lambda_0$, it is still in the form domain $H^1(\Lambda_0)$ of
the Neumann operator $-\Delta + q_a(x)$ on $\Lambda_0$. Therefore,
minimizing the right hand side of equation
(\ref{eqn:layerproperty3}) over all normalized functions in
$H^1(\Lambda_0)$ it is clear $E_0(0) \geq E_0(a)$.

To show that indeed $E_0(0)$ is strictly greater than $E_0(a)$ it
suffices to show that, when restricted to $\Lambda_0$, $\Psi_{a}$ is
not equal to a multiple of the Neumann ground state eigenfunction
corresponding to the potential $q_a$ on $\Lambda_0$. Suppose, for
contradiction, that $\Psi_{a}$ was such a multiple. Then by
construction the box $B = (-\frac{1}{2},-\frac{1}{2}+a) \times
(-\frac{1}{2},\frac{1}{2})^{d-1}$ is disjoint from the support of
the potential and $\Psi_{a}$ satisfies the equation
\begin{eqnarray}
-\Delta \Psi_{a} = E_0(0)\Psi_{a}
\end{eqnarray}
with Neumann conditions on the boundary of $B$. As $\Psi_{a}>0$, it
is the ground state of the Neumann problem on $B$. Thus $E_0(0)=0$
and $\Psi_{a}$ must be constant on $B$. This entails that $\Psi_{a}$
is harmonic outside the support of the potential, and since it is
constant on an open subset, by unique continuation of harmonic
functions it must be constant everywhere outside the support of the
potential. However this implies that alternative (ii) must hold, a
contradiction. Thus $E_0(0) > E_0(a)$.
\end{proof}

\begin{lemma}
For any positive integer $n$ and $a \in (0,\frac{1}{2}-r]$, $E_0(na)
> E_0((n+1)a)$ so long as $(n+1)a$ is less than or equal
to $\frac{1}{2}-r$. \label{lem:lem2}
\end{lemma}
%where $\textbf{$e_1$} = (1,0,\ldots,0)$%
\begin{proof}
To keep notations simple, we first show this for $n=1$. Again
consider the tube $L := \{x : |x_i| < \frac{1}{2},
i=2,\ldots,d$\}. Fix $a$ with $0<2a\le \frac{1}{2}-r$ and consider
a $2$-periodic extension, $w_{per}$, of the potential
\begin{equation} \label{defWper}
w(x) := q_a(x) + q_{-a+1}(x)
\end{equation}
on $L$ given by
\[
w_{per}(x) := \sum_{i\in \Z} w_{2i}(x).
\]
\begin{figure}[h]
  \centering
  \includegraphics[width=0.40\textwidth]{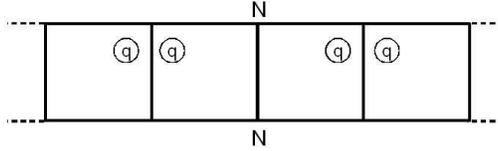}
  \caption{Periodic extension, $W_{per}(x)$, of
  $w(x) :=
w_{a}(x) + w_{-a+1}(x)$} \label{figperiodtwo}
\end{figure}
As before we consider the Neumann problem on $L$ for $-\Delta +
w_{per}(x)$, see Figure~\ref{figperiodtwo}, and let $\psi$
denote the Neumann ground state of $-\Delta +
q_{a}(x)$ on the unit cube $\Lambda_0$, normalized
to $\|\psi\|^2=\frac{1}{2}$. Construct a new function $\Psi$ on
all of $L$ by $2$-periodically extending
$\psi(x_1,x_2,\ldots,x_d)+\psi(-x_1+1,x_2,\ldots,x_d)$.

Symmetry and periodicity of the potential implies that $\Psi$ is a
smooth solution of
\[
-\Delta \Psi(x) + w_{per}(x)\Psi(x) = E_0(a) \Psi(x) \label{eqn:perlayer}
\]
on all of $L$. Now we proceed in analogy to
(\ref{eqn:layerproperty2}) and (\ref{eqn:layerproperty3}), this time
considering cells of length $2$ instead of unit cells and again
shifting by $a$ to the right. We find
\begin{eqnarray}
E_0(a) & = & \int_{\{x\in L:
-\frac{1}{2}-a<x_1<\frac{3}{2}-a\}} \left(|\nabla
\Psi(x)|^2 +
w_{per}(x) \Psi(x)^2\right) dx \nonumber \\
 &=& \int_{\{x\in L:
-\frac{1}{2}<x_1<\frac{3}{2}\}} \left(|\nabla
\Psi_{a}(x)|^2 + (q_{2a}(x)+q_{1}(x))
\Psi_{a}^2(x)\right) dx. \label{eqn:perlayer3}
\end{eqnarray}
As above we conclude that $E_0(a)$ is not smaller than the first
Neumann eigenvalue $\tilde{E}$ of $-\Delta + q_{2a}+q_{1}$ on
$\{x\in L: -\frac{1}{2}<x_1<\frac{3}{2}\}$. A further application of
the argument in the last paragraph of the proof of
Lemma~\ref{lem:lem1} shows that $\Psi_a$ restricted to the unit 2
cell $\{x \in L : -\frac{1}{2} < x_1 < \frac{3}{2} \}$ is not a
multiple of the ground state eigenfunction corresponding to
$\tilde{E}$. Thus $E_0(a)$ is strictly greater than $\tilde{E}$.

Imposing an additional Neumann condition at $x_1=\frac{1}{2}$ can
not increase the lowest eigenvalue, thus
\[
E_0(a) > \tilde{E} \geq \min \{ E_0(2a), E_0(0)\} = E_0(2a)
\]
by Lemma~\ref{lem:lem1}, which concludes the proof for $n=1$. The
crucial idea which allowed us to reduce the $n=1$ claim to
Lemma~\ref{lem:lem1} was that the term $q_{-a+1}$ in (\ref{defWper})
was shifted back into the center of the cube $\{x\in L:
\frac{1}{2}<x_1<\frac{3}{2}\}$ in (\ref{eqn:perlayer3}). The same
mechanism can now be used to inductively prove the claim for all
$n$.
\end{proof}

We can now readily complete the proof of \eqref{eq:Emon}: As $a$
above was arbitrary, Lemma~\ref{lem:lem2} implies that $E_0(a)$ is
strictly decreasing on the set of all dyadic numbers in
$[0,\frac{1}{2}-r]$. As this set is dense in $[0,\frac{1}{2}-r]$
and $E_0(a)$ is continuous, then $E_0(a)$ is strictly decreasing
on all of $[0,\frac{1}{2}-r]$. Therefore assuming (ii) is false, (i) must be true.

\section{Proof of Theorem~\ref{thm:neumannconvex}}
\label{sec:neumannconvex}

In Theorem~\ref{thm:neumannconvex} it is assumed that the domain
and the potential are smooth. Thus we have by elliptic regularity
that $C^\infty(\overline{D})$ is a form core for the Neumann
operator $H_D^N(a)=-\Delta_D^N+q_a$. Moreover, the eigenfunctions
are all in $C^\infty(\overline{D})$ and have normal derivative
zero on $\partial D$.

We call the eigenvalues $E_k(a)$, ordered and accounting for
multiplicity, and the normalized eigenfunctions $u_k(x;a)$,
$k=0,1,\ldots$. We choose the ground state $u_0$ strictly positive
and all other $u_k$ real. They form an orthonormal basis in
$L^2(D)$. Thus, for any function $f \in H^1(D)$ we have that
\begin{equation} \label{form}
\sum_{k=0}^\infty E_k (u_k, f)^2 = \int_D \left[|\nabla f(x)|^2+
q_a(x) f(x)^2\right] dx \ ,
\end{equation}
where $(\cdot,\cdot)$ denotes the inner product on $L^2(D)$. Note
that $f$ is not in the domain of the operator, just in the form
domain. See \cite{Barry4} and \cite{Evans}.

Let $N$ denote the outward normal vector field on $\partial D$. It
will be convenient to use that $N$ can be extended to a smooth
vector field in a neighborhood of $\partial D$. To see this, first
work in a neighborhood of a fixed point of the surface. Without
loss (i.e.\ up to a rigid motion) we can choose this point to be
the origin and the surface to be given by
\begin{equation} \label{eq:local1}
x_d = f(x_1,\ldots,x_{d-1})
\end{equation}
in a vicinity of the origin, where $f$ is smooth,
\begin{equation} \label{eq:local2}
f(0,\ldots,0)=0 \quad \mbox{and} \quad \nabla f(0,\ldots,0)=0 \ .
\end{equation}
Thus at a point $p=(x_1,\ldots,x_{d-1},f(x_1,\ldots,x_{d-1}))$ near
$0$ one has
\[ N(p) = \frac{(-\nabla f(x_1,\ldots,x_{d-1}),1)}{|(-\nabla
f(x_1,\ldots,x_{d-1}),1)|} \ .
\]
This can be extended smoothly to $x=(x_1,\ldots,x_{d-1},x_d)$ near
$0$ by
\[
N(x) = N(x_1,\ldots,x_{d-1},f(x_1,\ldots,x_{d-1})).
\]
We get a
global extension of $N$ to a neighborhood of $\partial D$ by using
compactness of $\partial D$ and a standard partition of unity
argument.

In this neighborhood we define the matrix-valued smooth vector
field $K = (K_{ij}) = (\partial_i N_j)$. The restriction of $K$ to
$\partial D$ is the curvature matrix of the surface. Indeed, in
the local coordinates used above, we have
\[ \partial_i N_j(0) = \left\{ \begin{array}{ll} -(\partial_i
\partial_j f)(0) & \mbox{if $1\le i,j \le d-1$}, \\ 0 & \mbox{if
$i=d$ or $j=d$}. \end{array} \right.
\]
We have assumed that $D$ is strictly convex. This means that the
Hessian of $f$ is negative definite and thus at every $p\in
\partial D$ the restriction of $K(p)$ to the tangent plane at $p$
is positive definite.

The following identity for $E_0(a)$ is the main technical
ingredient into our proof. Here $\Delta E_0$ and $\nabla E_0$
refer to the $a$-derivatives of $E_0$. Otherwise, all symbols such
as $\partial_i$, $\nabla$, $\Delta$ denote derivatives with
respect to the spatial variable. $B(\cdot,\cdot)$ is the bilinear
form
\[ B(u,v) = (u,\Delta v) - (\Delta u,v).\]

\begin{lemma}[\bf{Second order perturbation theory}] \label{perturbation}
The ground state energy satisfies the equation
\begin{equation} \label{eq:bigone}
\Delta E_0 - 4(u_0, \nabla u_0) \nabla E_0 = -2 \int_{\partial D}
\nabla u_0 \cdot K \nabla u_0\,dS
 -  2\sum_{k \not= 0} \frac{\sum_i B(u_k, \partial_i
u_0)^2}{E_k-E_0} \ .
\end{equation}
\end{lemma}

\begin{proof} We start with the second order perturbation theory formula
\begin{equation} \label{eq:2ndorderpert}
\Delta  E_0 = (u_0, (\Delta q_a) u_0) -2\sum_{k \not= 0}
\frac{\sum_i (u_0,(\partial_i q_a) u_k)^2}{E_k-E_0},
\end{equation}
see Section~\ref{subsec:pert}. Differentiating the eigenvalue
equation yields
\[
(\nabla q_a) u_0 = E_0 \nabla u_0 -(-\Delta+ q_a) \nabla u_0
\]
and therefore
\begin{equation} \label{workhorse}
(u_k, (\nabla q_a) u_0) =-(E_k-E_0)(u_k, \nabla u_0) +B(u_k,
\nabla u_0) \ .
\end{equation}
Hence
\begin{eqnarray*}
\Delta E_0 &=& (u_0, (\Delta q_a) u_0) - 2\sum_{k \not= 0}
\left[(E_k-E_0)(\nabla u_0, u_k)^2 -2(\nabla u_0, u_k)\cdot B(u_k, \nabla u_0) \right] \\
& & \mbox{} - 2\sum_{k \not= 0} \frac{\sum_i B(u_k, \partial_i
u_0)^2}{E_k-E_0}
\end{eqnarray*}
which, using (\ref{workhorse}) once more, can be rewritten as
\begin{eqnarray*}
\Delta E_0 &=& (u_0, (\Delta q_a) u_0) + 2\sum_{k \not= 0}[B(u_k,
\nabla u_0)+(u_k, (\nabla q_a) u_0)]\cdot (u_k,\nabla u_0) \\
& &\mbox{}- 2\sum_{k \not= 0} \frac{\sum_i B(u_k, \partial_i
u_0)^2}{E_k-E_0} \\
&=& (u_0, (\Delta q_a) u_0) + 2\sum_{k}[B(u_k, \nabla u_0)+(u_k, (\nabla q_a) u_0)]\cdot (u_k,\nabla u_0)\\
& & \mbox{}-2 [B(u_0, \nabla u_0)+(u_0, (\nabla q_a) u_0)]\cdot (u_0,\nabla u_0)\\
& & \mbox{} - 2\sum_{k \not= 0} \frac{\sum_i B(u_k, \partial_i
u_0)^2}{E_k-E_0} \\
&=& (u_0, (\Delta q_a) u_0)+ 2 (\nabla u_0, (\nabla
q_a) u_0)+2\sum_{k}B(u_k, \nabla u_0)\cdot (u_k,\nabla u_0)\\
& & \mbox{} - 2 [B(u_0, \nabla u_0)+(u_0,(\nabla q_a) u_0)]\cdot (u_0,\nabla u_0)\\
& & \mbox{} - 2\sum_{k \not= 0} \frac{\sum_i B(u_k, \partial_i
u_0)^2}{E_k-E_0} \ ,
\end{eqnarray*}
where finally the completeness relation of the $u_k$ was used. It
is clear that
\[
(u_0, (\Delta q_a) u_0) + 2(\nabla u_0, (\nabla q_a) u_0)  = 0 \
\]
and using again (\ref{workhorse}) with $k=0$ we can simplify further
and get
\begin{eqnarray} \label{eq:E0eq}
\Delta E_0 &=&  2\sum_{k} B(u_k, \nabla u_0) \cdot (u_k,\nabla u_0) -
4(u_0, (\nabla q_a) u_0)\cdot (u_0,\nabla u_0) \nonumber\\
& & \mbox{}- 2\sum_{k \not= 0} \frac{\sum_i B(u_k, \partial_i
u_0)^2}{E_k-E_0} \ .
\end{eqnarray}
Recall that $B(u,v)=(u,\Delta v) -(\Delta v, u)$ and hence
\begin{equation} \label{eq:defofB}
\sum_{k} B(u_k, \nabla u_0) \cdot (u_k,\nabla u_0) =\sum_{k}[(u_k,
\Delta \nabla u_0) - (\Delta u_k, \nabla u_0)]\cdot (u_k,\nabla
u_0) \ .
\end{equation}
Since $u_0 \in C^\infty(\overline{D})$ we know that the vector
$\Delta \nabla u_0$ has square integrable components and hence
\[
\sum_{k}(u_k, \Delta \nabla u_0)\cdot (u_k,\nabla u_0) = (\nabla
u_0, \Delta \nabla u_0) \ .
\]
The second term of \eqref{eq:defofB} we write as
\[
\sum_{k}([-\Delta + q_a] u_k, \nabla u_0)\cdot (u_k,\nabla u_0)
-\sum_{k}(q_a u_k, \nabla u_0)\cdot (u_k,\nabla u_0) \ .
\]
The second sum equals $\sum_j (\partial_j u_0, q_a \partial_j
u_0)$ while the first sum is
\[
\sum_j \sum_{k}E_k ( u_k, \partial_j u_0)^2 =\sum_j \int_D [|\nabla
\partial_j u_0|^2 + q_a(x) (\partial_j u_0)^2] dx
\]
since $\partial_j u_0$ is in the form domain. Collecting terms we
find from \eqref{eq:defofB} that
\begin{eqnarray} \label{eq:Green}
\sum_{k}B(u_k, \nabla u_0) \cdot (u_k,\nabla u_0) & = &\sum_j
[(\partial_j u_0, \Delta \partial_j u_0) +\Vert \nabla \partial_j
u_0 \Vert^2] \nonumber \\
& = & \sum_j \int_{\partial D} (\partial_j u_0) N\cdot \nabla
(\partial_j u_0)\,dS,
\end{eqnarray}
where Green's identity was used. On an open neighborhood of
$\partial D$ we have
\begin{eqnarray} \label{eq:curvid}
\sum_j (\partial_j u_0) N\cdot \nabla (\partial_j u_0) & = &
\sum_j (\partial_j u_0) \partial_j N \cdot \nabla u_0 - \sum_{j,i}
(\partial_j u_0) (\partial_j N_i) (\partial_i u_0) \nonumber \\
& = & \nabla u_0 \cdot \nabla (N\cdot \nabla u_0) - \nabla u_0
\cdot K \nabla u_0 \ ,
\end{eqnarray}
where $K$ is the curvature matrix defined above. Using that
$N\cdot \nabla u_0 =0$ on $\partial D$ one has that the first term
is $\nabla_t u_0 \cdot \nabla_t (N\cdot \nabla u_0)$ for points in
$\partial D$, where $\nabla_t$ denotes the component of the
gradient in the tangential directions. However, $N\cdot \nabla
u_0=0$ and smoothness of $\partial D$ also implies $\nabla_t
(N\cdot \nabla u_0)=0$ and thus $\nabla u_0 \cdot \nabla (N\cdot
\nabla u_0) =0$ on $\partial D$. Thus \eqref{eq:Green} and
\eqref{eq:curvid} yield
\[
\sum_k B(u_k, \nabla u_0) \cdot (u_k, \nabla u_0) =
-\int_{\partial D} \nabla u_0 \cdot K \nabla u_0\,dS.
\]
After substituting this and the first order perturbation formula
\begin{equation} \label{eq:firstorderpert}
\nabla E_0 = -(u_0, (\nabla q_a) u_0)
\end{equation}
into \eqref{eq:E0eq} we arrive at \eqref{eq:bigone}.
\end{proof}

\begin{lemma}\label{symmetry}
Assume that $q$ is a smooth potential with compact support in $D$.
Assume that the Neumann ground state $u_0(x;a)$ for some fixed
$a_0 \in G$ is constant on $\partial D$. Then there exists an open
neighborhood of $a_0$ where $E_0(a_0) \ge E_0(a)$.
\end{lemma}
\begin{proof} By shifting coordinates we assume that $a_0=0$.
We shall proceed by a trial function argument. Consider the
problem with the shifted potential $q_a(x):=q(x-a)$. Denote by
$D_a=D+a$ the shifted domain, i.e., the function
$u_a(x):=u_0(x-a;0)$ solves $-\Delta u_a + q_a u_a = E_0(0) u_a$,
with a Neumann boundary condition on $\partial D_a$ as well as
being constant on $\partial D_a$. We shall construct a trial
function $\phi$ in the following fashion. In the intersection of
$D_a$ with $D$ we set $\phi = u_a$ and in $D \backslash D_a$ we
set $\phi$ to be a constant which equals the boundary value of
$u_a$. Note that $\phi \in H^1(D)$. By the variational principle
\[
E_0(a) \le \frac {\int_D |\nabla \phi|^2 +q_a \phi^2 dx}{\int_D
\phi^2 dx} \ .
\]
For $a\in G$ the right side equals
\begin{equation} \label{inequality}
\frac {\int_{D\cap D_a} |\nabla u_a|^2 +q_a u_a^2 dx}{\int_D
\phi^2 dx} \le \frac {\int_{D_a} |\nabla u_a|^2 +q_a u_a^2
dx}{\int_D \phi^2 dx} = E_0(0)  \frac{\int_D u_0^2 dx}{\int_D
\phi^2 dx} \ .
\end{equation}
In the case $E_0(0)=0$, this implies $E_0(a)\le 0$, as was to be
shown.

Next we claim that for $a$ sufficiently small
\[
\frac{\int_D u_0^2 dx}{\int_D \phi^2 dx} < 1
\]
if $E_0(0) > 0$ and
\[
\frac{\int_D u_0^2 dx}{\int_D \phi^2 dx} > 1
\]
if $E_0(0) < 0$. This yields the lemma for the remaining cases. If
we denote by $c$ the boundary value of $u_0$ we find that the
claim follows once we show that in a vicinity of the boundary,
$u_0 < c$ for $E_0(0) > 0$ and $u_0 > c$ for $E_0(0) < 0$. To see
this, fix a point on the boundary, call it the origin and use the
local coordinates \eqref{eq:local1} and \eqref{eq:local2} above.
The normal vector at $0$ is $(0, \dots, 0,1)$ and hence the normal
derivative equals $\partial_d u_0(0) = 0$. Further, since $u_0$ is
constant on the boundary we find by differentiating $u_0(x_1,
\dots, x_{d-1}, f(x_1,\dots ,x_{d-1})) \equiv c$ that
\[
\partial_i \partial_j u_0(0) = -\partial_d u_0(0) \partial_i \partial_jf(0) = 0
\]
for $i,j =1, \dots d-1$. Hence
\[
\partial_d \partial_d u_0(0) =\Delta u_0(0) = -E_0(0) u_0(0)
\]
which is negative for $E_0(0) > 0$ and positive for $E_0(0) < 0$.
As this holds at all points of the boundary, we get the required
property of $u_0$ in a vicinity of the boundary.
\end{proof}

We remark that in the case $E_0(a_0)\not= 0$, the above proof
actually gives the strict inequality $E_0(a_0)>E_0(a)$ for $a$
close to $a_0$, i.e.\ $E_0(a_0)$ is a strict local maximum. This
is the case, for example, if $q$ is sign-definite, and allows for
a shorter argument in the following completion of the proof of
Theorem~\ref{thm:neumannconvex}.

\begin{proof}[Proof of Theorem~\ref{thm:neumannconvex}]
Assume that $E_0$ attains its minimum value in $G$ say at the point
$a_0$. This entails that $\nabla E_0(a_0)=0$ and $\Delta E_0(a_0)
\ge 0$. Using the Lemma \ref{perturbation} we find that
\[
\Delta E_0 (a_0) = -2 \int_{\partial D} \nabla u_0 \cdot K \nabla
u_0\, dS - 2\sum_{k \not= 0} \frac{\sum_i B(u_k,
\partial_i u_0)^2}{E_k-E_0} \ .
\]
The right side is non-positive, since $D$ is convex. It cannot be
strictly negative, since that would contradict the assumption that
$E_0(a)$ has a local minimum at $a_0$.

Thus the right side must vanish. Since $D$ is strictly convex, the
first term can vanish only if $u_0$ is constant on the boundary
(recall that $K$ is positive definite on the tangent space at each
point of $\partial D$ and that $\nabla u_0$ is a tangent vector).
However, Lemma \ref{symmetry} shows that $E_0(a_0) \ge E_0(a)$ in a
neighborhood of $a_0$. Thus $E_0(a)=E_0(a_0)$ in this neighborhood.
Since the set where $E_0(a_0) = E_0(a)$ is closed and open in $G$
(the above argument applies to every $a$ with $E_0(a_0)=E_0(a)$) the
function $E_0(a)$ must be constant. Assuming now that $E_0(a)$ is
constant, $u_0$ is constant on the boundary and moreover, there must
be equality in (\ref{inequality}). This means that
\[
\int_{D_a \backslash D} |\nabla u_a|^2 dx = 0 \ .
\]
Hence $u_a$ must be constant in $D_a \backslash D$ and therefore,
for small $a$, $0 = -\Delta u_a = E_0(a)u_a$ there. Thus, $E_0(a)
\equiv 0$. Since the support of $q_a$ is a subset of $D$ (as $a\in
G$), $u_a$ is constant in the non-empty open set $D_a \backslash
\overline{D}$, which is disjoint from the support of $q_a$. Since it
is harmonic outside the support of $q_a$ it must be constant there
too. It follows that $u_0$ is constant outside the support of $q$.
\end{proof}

\section{Discussion: Extensions and open problems}

We conclude with some remarks about possible generalizations and
open problems related to our main results.

(i) Theorem~\ref{thm:neumannsquare} and its proof immediately
generalize to the Neumann problem on an arbitrary rectangular box
$\{x: |x_i|<\ell_i, i=1,\ldots,d\}$ rather than the unit cube
$\Lambda_0$, which we chose to keep notations simple. This also
gives a corresponding version of Theorem~\ref{thm:main}, where
$\Z^d$ is replaced by an arbitrary rectangular lattice.

(ii) In Theorems~\ref{thm:main} and \ref{thm:neumannsquare} we
also may replace the obstacle $q$ by an reflection symmetric hole
with Dirichlet boundary conditions, often interpreted as an
infinite barrier. More precisely, let $C\subset \{x:|x_i|<r,
i=1,\ldots,d\}$ be closed and reflection symmetric. Let $C_{a} =
C+a$ and $H^N_{a} := -\Delta$ on $\Lambda\setminus C_{a}$ with
Neumann conditions on $\partial \Lambda$ and Dirichlet conditions
on $\partial C_{a}$. Then $E_0(a)$ is minimized when the hole is
in a corner of $\Lambda_0$ and $\inf \sigma(H(\omega))$ is
minimized for a periodic configuration of $2^d$-clusters of holes.
While we expect that Theorem~\ref{thm:neumannconvex} also extends
to this situation, at least for holes with smooth boundary, our
proof does not extend directly.

(iii) Theorem~\ref{thm:neumannconvex} covers the situation of a
radially symmetric potential (or Dirichlet hole) placed in a
spherical domain, where all placements of $q$ which touch the
boundary are equivalent minimizing positions. However, in this case
the methods of Lemma~\ref{symmetry} may be used to also show that
the maximal position occurs when the potential is centered in the
domain.

To see this, suppose $D$ is a spherical domain centered at 0 and let
$q$ be a radially symmetric potential compactly supported in $D$,
also centered at 0. Radial symmetry of the domain and potential then
imply the ground state eigenfunction, $u_0$, corresponding to
$E_0(0)$ is radially symmetric. Thus it satisfies the conditions of
Lemma~\ref{symmetry}. If $E_0(0) \geq 0$ one may show by reducing
the problem to one dimension, or by using maximum principles, that
outside the support of the potential $u_0 \leq c$, where $c$ denotes
the value of $u_0$ on the boundary of $D$. Similarly if $E_0(0) \leq
0$, $u_0 \geq c$ outside the support of the potential. One may
conclude using the arguments of Lemma~\ref{symmetry} that $E_0(0)
\geq E_0(a)$ for every $a \in \overline{G}$. This then leads to the
following alternative for the case of a spherical domain centered at
0 with radially symmetric potential: either $E_0(a) \equiv 0$, in
which case the corresponding eigenfunction is constant outside the
support of the potential \emph{or} $E_0(a)$ assumes a strict maximum
at $E_0(0)$ and strict minima when the support of the potential
touches the boundary.

(iv) Theorems~\ref{thm:neumannsquare} and \ref{thm:neumannconvex}
are proven with very different methods and apply to mutually
exclusive classes of domains (rectangular boxes vs strictly convex
domains). It would be desirable to find a method of proof which
covers both results, as this method would most likely also cover
more general polygons and polyhedra. Particularly interesting
cases would be equilateral triangles or hexagons, as they tile the
plane and would lead to a corresponding extension of
Theorem~\ref{thm:main}.

(v) One can view Theorem~\ref{thm:main} as a mechanism in which
the nuclei of a solid self-organize into a simple periodic
pattern, given a density condition (exactly one site per cube). It
would be wrong, in our opinion, to see this as a model for
crystallization since the regularity of the pattern is to a large
extend determined by the density condition.  Real crystallization,
however, cannot be explained by the interaction of one electron
with nuclei alone. It is a many-body effect and the nuclear
repulsion and, more importantly, the Pauli exclusion principle
play a role. Further, one needs sufficiently many electrons, e.g.,
a half filled band. Indeed, there have been results in this
direction in \cite{KennedyLieb} for the Falicov-Kimball model, a
variant of the Hubbard model where the nuclei are treated
classically and sit on a lattice where the electrons hop.
Crystallization was shown in \cite{KennedyLieb} for the half
filled band. In our model, without the density condition, we
expect that the nuclei would stick together.  While this is an
open question there is some evidence in this direction. For bosons
it was shown in \cite{KennedyLieb} that the nuclei indeed stick
together.

It would be interesting to consider an extension of our model, a
continuous analog of the Falicov-Kimball model, in which one
considers a finite periodic array of cubes on a torus. Assuming
the same number of spinless fermions as the number of cubes and
assuming one nucleus in each cube, it is not unreasonable to
expect that in an energy minimizing configuration the nuclei sit
at the center of each cube. This is an interesting open question.
Needless to say that the methods in this paper have no bearing on
this problem. For an overview of the Falicov-Kimball model the
reader may consult \cite{FreLieUel}.

\vspace{.5cm}

\noindent {\bf Acknowledgements:} The authors are indebted to Jean
Bellissard for many useful discussions and suggestions which
substantially improved this work. G.~S. would also like to thank
Michael Levitin, from whom he originally learned the ideas used in
Section~\ref{sec:clustering}. He also acknowledges hospitality and
support at the Isaac Newton Institute of the University of
Cambridge, where part of this work was done, as well as partial
support through NSF grant DMS~0245210. M.~L. would like to
acknowledge partial support through NSF grant DMS-0600037.

\end{document}